\documentclass[10pt,twocolumn]{article}

\usepackage{multicol}
\usepackage{graphicx}
\usepackage{amsmath}
\usepackage{lipsum}
\usepackage{color}
\usepackage{siunitx}
\usepackage{authblk}
\usepackage{subfigure}
\usepackage[margin=2cm]{geometry}

\title{Reorganization of a granular medium around a localized transformation}
\author[1]{Aymeric Merceron}
\author[1]{Alban Sauret}
\author[1]{Pierre Jop}
\affil[1]{Surface du Verre et Interfaces, UMR 125 CNRS/Saint-Gobain, 39, quai Lucien Lefranc, F-93303 Aubervilliers, Cedex, France}

\date{ }
\begin{document}

\twocolumn[
    \begin{@twocolumnfalse}
        \maketitle
        \begin{abstract}
            Physical and chemical transformation processes in reactive granular media involve the reorganization of the
structure. In this paper, we study experimentally the rearrangements of a two-dimensional (2D) granular packing
undergoing a localized transformation. We track the position and evolution of all the disks that constitute the
granular packing when either a large intruder shrinks in size or is pulled out of the granular structure. In the two
situations the displacements at long time are similar to 2D quasistatic silo flows whereas the short-time dynamic is
heterogeneous in both space and time. We observe an avalanchelike behavior with power-law distributed events
uncorrelated in time. In addition, the instantaneous evolutions of the local solid fraction exhibit self-similar
distributions. The averages and the standard deviations of the solid fraction variations can be rescaled, suggesting
a single mechanism of rearrangement. \\
            \medskip\medskip\medskip
        \end{abstract}
    \end{@twocolumnfalse}
]

\section{Introduction}

Sintering, glass melting and other industrially relevant processes turn large assemblies of grains into homogeneous final products. Such processes are often  complex as they can involve both chemical and physical transformations of the granular packing leading to the rearrangement of the structure of the material. For instance, dilatations of the material and phase transitions affect the volume of the particles \cite{turnbull_scaling_2011,ludewig_bernal_2015}. Consequently, the shapes of particles are modified \cite{kintea_shape_2015} as well as their flowing properties through a lubrication film \cite{xu_lubrication_2007} or the formation of capillary bridges \cite{herminghaus_dynamics_2005}. In addition, chemical reactions can also modify the surface and nature of each individual grain \cite{gouillart_situ_2012}. These different phenomena occur at the same time and lead to the final homogeneous product through complex paths. Modeling the behavior of such systems involves different fields of research. Typically, mechanics and statistical physics are used to predict the behavior of granular assemblies under flow, rearrangements or perturbations \cite{andreotti_granular_2013}. The rich behavior of granular media has led to extensive studies considering their rheology \cite{jop_constitutive_2006}, the criteria of mechanical stability \cite{wyart_stability_2012}, and the properties of wet granular assemblies \cite{mitarai_wet_2006}. Other fields of research have been considered to characterize a granular assembly in more complex situations. For instance, the behavior of reactive powders is modeled in chemical engineering \cite{lame_situ_2003} and rock transformations have been characterized in geology by focusing on thermodynamics of reactions \cite{wong_conductivity_1984}. 


However, despite these different approaches, the reorganization of a granular assembly at short time scales remains poorly understood. In particular, few studies have considered a granular medium whose particles are undergoing reactive transformations. Some recent studies have characterized the rearrangement of a reactive granular packing, often icy particles than can melt \cite{turnbull_scaling_2011,dorbolo_how_2012}, but experimental investigations remain rare, likely because of the difficulties in monitoring the location of the individual grains as well as their properties. Therefore, most studies have focused on granular flows around a unique intruder using mechanical \cite{kolb_rigid_2013,coulais_shear_2014,hamm_laser_2006}, rheological \cite{seguin_local_2016}, or purely kinematic approaches \cite{seguin_dense_2011,kolb_flow_2014}. In addition, these different studies were often performed using an intruder penetrating the granular packing. Flows of granular materials induced by gravity in silos and hoppers were also largely considered \cite{choi_velocity_2005} and mainly deal with mean field flows of the granular material over a long time. But such flows induce significant reorganizations of the granular packing leading to jamming problems \cite{janda_jamming_2008,zuriguel_silo_2011}. Similar silo problems have been considered in mining applications where grains are also extracted from a bottom orifice under gravity at very low flow rate \cite{castro_study_2007,melo_kinematic_2008}. However, all these studies only characterized and modeled the mean displacement fields of the granular assembly but did not characterize the instantaneous response of the packing and the distribution of the size of events during the flowing process. The rearrangements at small time scales and length scales remain poorly explored although they can govern the evolution of a reactive granular packing. 


In this paper, we experimentally investigate the quasistatic response of a two-dimensional (2D) granular assembly undergoing a localized transformation. We model the transformation using two methods: (i) with a single large grain that shrinks in size and (ii) with the slow extraction of a large intruder in the granular assembly. The position of each grain during the rearrangement process is accurately characterized. This method also allows us to observe the rearrangements of the granular assembly over large spatial and time scales. We thus quantify the typical amplitude of a rearrangement -
the fraction of the granular assembly that is involved in the localized transformation. We also determine how the
granular structure is modified during the reorganization. We present our experimental methods in Sec. \ref{section:experiments}. We qualitatively describe the long-term displacements using a melting grain and a mechanical intruder in Sec. \ref{section:long_term_dpts}. We then rely on the experiments performed with a mechanical intruder to characterize the statistics of reorganization events. These quantitative results are reported in Sec. \ref{section:short_term_dpts} where we describe the reorganization process in term of both the dynamics of displacements and instantaneous structural evolutions. Finally, the results are discussed in Sec.~\ref{section:discussion}.


\section{Experiments} \label{section:experiments}

\begin{figure}
\centering
\includegraphics{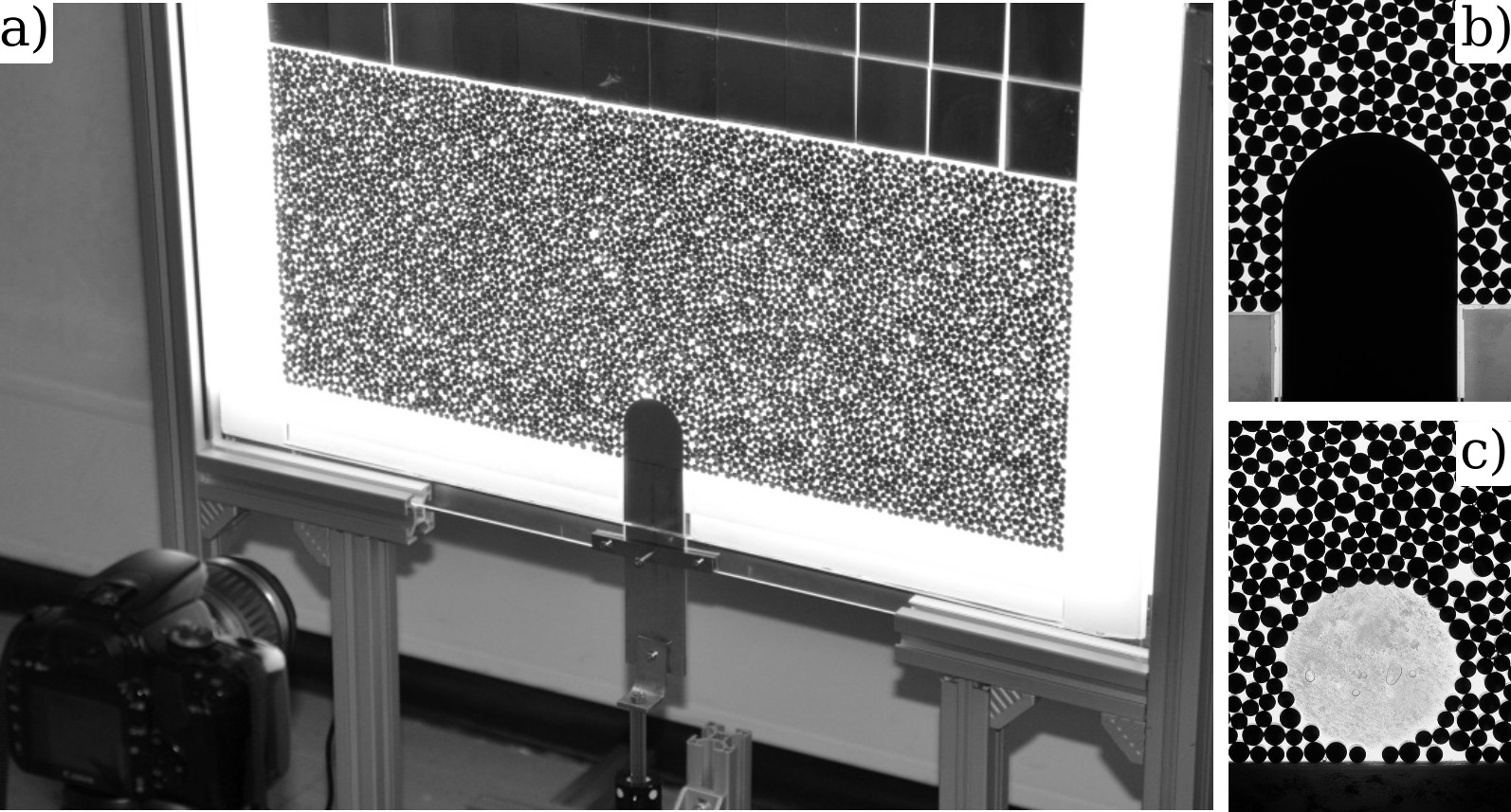}
\caption{(a) Picture of the experimental setup showing the mechanical intruder and the 2D packing. (b) Zoom on the intruder used in the mechanical-intruder experiment and on (c) the initial DMSO disk used in the melting block experiment. Pictures in both (b) and (c) are 6.5 cm wide.}
\label{fig:setup}
\end{figure}

The reorganization of a granular medium around a localized transformation is studied using a two-dimensional system, shown in Figs.~\ref{fig:setup}(a)-(c), which allows us to track the time evolution of the position of each individual grain. We consider two types of localized transformation: the shrinkage of a large grain and the extraction of a large intruder. The first method is called the melting block experiment and the second one the mechanical-intruder experiment. In both cases, we use a two-dimensional cell made of two parallel glass plates (30 cm high and 30 cm wide for the melting block experiment or 60 cm wide for the mechanical-intruder experiment) separated by a gap of 3.1~mm. The cell is filled with 5000 or 3000 (depending on the width of the cell) bidisperse rigid disks 4 and 5 mm in diameter (proportion 10:7) to avoid crystallization. In the following all lengths are expressed in terms of the diameter of smallest grains, $d_g=4\,{\rm mm}$.   

%
%
 

In a melting block experiment, a pure dimethylsulfoxide (DMSO) disk 40 mm in diameter (10 $d_g$) and 3~mm thick is placed at the bottom of the granular assembly and surrounded by inert metal disks [Fig.~\ref{fig:setup}(c)]. To prevent the formation of capillary bridges during the melting process and to provide a better thermal homogeneity, the entire granular packing is immersed into a mixture of DMSO and water (20$\%$ v/v) initially thermalized at $-5\, ^o$C. The melting point of the DMSO is $18.5\, ^o$C; therefore, the reactive grain slowly melts while the system warms up to room temperature, leading to the reorganization process. Hydrodynamic interactions are present but they remain negligible compared to the weight of the disks since the density of the metal disk is $\rho_{g} = 8$~g\,cm$^{-3}$ and the dynamic viscosity of the water/DMSO mixture is $\eta_{mixture}\approx 8~$mPa s. The typical time scale of the melting of the DMSO disk is about 40 min, which ensures a quasistatic evolution of the surrounding granular assembly. However, the control of these experiments is very sensitive as some heterogeneities can be present in the DMSO disk leading to variations in the melting process. In addition, this method, although the closest to a reactive transformation, makes it difficult to conduct a large quantity of experiments. To overcome the challenges, we use an alternative method, the mechanical-intruder experiment, that also produces a quasistatic reorganization of the granular assembly.


In a mechanical-intruder experiment, a semicircular metallic intruder 40 mm in diameter (10 $d_g$) and 3~mm thick is inserted at the bottom of the cell and is surrounded by inert disks [Figs.~\ref{fig:setup}(a) and \ref{fig:setup}(b)]. To mimic a decrease of volume similar to the melting grain, the intruder is slowly pulled down out of the cell at a constant speed of 0.05 mm s$^{-1}$ using a linear translation stage. Similarly to the melting block experiment, the granular assembly follows a quasistatic evolution. In this situation, the granular packing remains perfectly dry in contrast with the melting block experiment so that grains only interact through friction.


In both situations, the experimental protocol is the following. First, the inert grains are mixed and horizontally inserted in the 2D cell at the same moment as the intruder to ensure that no preloading forces are exerted close to the intruder which could lead to an inhomogeneity in the packing. The method ensures that the granular packing forms a dense and disordered assembly at the beginning of the experiment. Metallic weights are also set horizontally. Then, the cell is tilted into a vertical position so that the system is driven by gravity. Metallic weights at the top ensure a uniform load all along the top of the granular assembly throughout the experiment.  Because of the low friction forces compared to the weight of the disks, the granular packing instantaneously reaches an initial stable state. Therefore, the rearrangements that may appear at the beginning of an experiment are not due to initial fragile states. Finally, depending on the experiment, the intruder is attached to the linear translation stage and then pulled down or the packing is immersed in the mixture mentioned above and the disk starts to melt, which defines the time $t=0$ in our experiments. A light-emitting diode panel is set behind the granular assembly and a high-resolution camera (Nikon D7000) records the time evolution of the packing at time intervals $\delta t=10\,{\rm s}$ for the melting block experiment and $\delta t=5\,{\rm s}$ for the mechanical-intruder experiment with a resolution of 4928 $\times$ 3264 pixels. For the mechanical intruder, the vertical displacement of the intruder between two snapshots is equal to $0.25\,{\rm mm} =d_{g}/16$. The pictures recorded during the experiments are processed with a segmentation protocol \cite{zhao_length_2013} to obtain a subpixel accuracy (inferior to $33\,\mu{\rm m}$) of the position of the disks. Then, tracking techniques are used to measure the short-term and long-term displacements. We have also performed tessellation methods to analyze the structural evolutions at different spatial scales.  It is worth noting that the length scale of both the translation of the mechanical intruder and the shrinking intruder between two snapshots is far above the resolution of the image processing. To obtain reliable statistics and compare the two situations, we have performed 20 times the melting block experiment and 100 times the mechanical-intruder experiment. 


We first present the phenomenological behavior observed in the experiments. As the intruder is melting or as it is being pulled down, we observe that the response of the granular packing is heterogeneous in space and time. For instance, although some reorganization events involve only few grains, large events mobilize almost the entire granular packing. In addition, the trajectories of some grains can be relatively steady in time whereas strongly intermittent behavior can be found elsewhere in the packing. Moreover, local structural variations, {\textit{i.e.}}, grain density, can be observed through compacting and dilating zones that are not necessarily close to the intruder.


\section{Long term displacements} \label{section:long_term_dpts}

\begin{figure*}
\centering
\includegraphics[width=\textwidth]{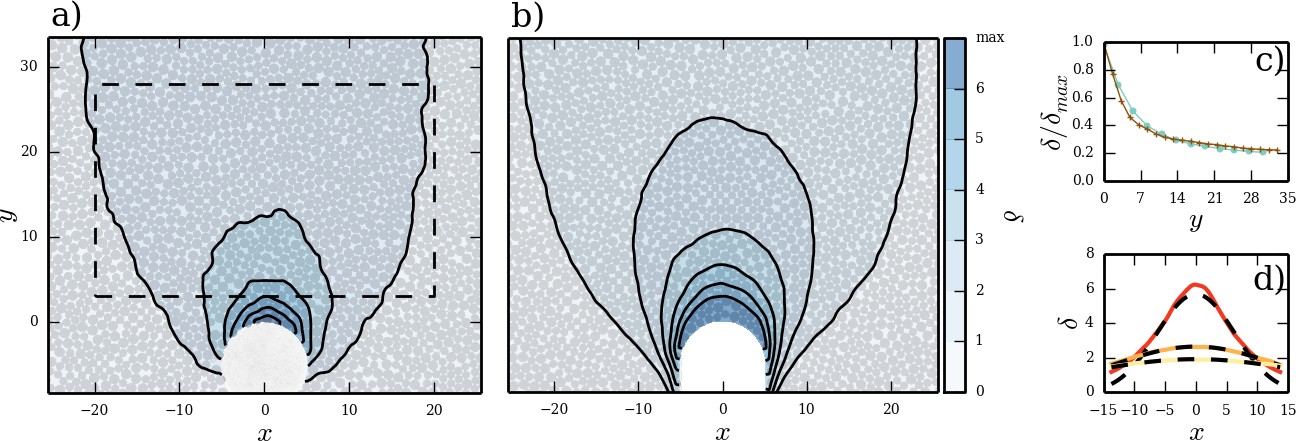}
\caption{Vertical displacement field between the initial and the final state for (a) the melting block setup (averaged over 20 experiments) and (b) the mechanical-intruder setup (averaged over 100 experiments). (c) Vertical normalized displacement field obtained at $x$=0 for the melting block (brown crosses) and mechanical-intruder (green circles) experiments. (d) Horizontal normalized displacement field extracted at $y$= 2.7, 13.6, 27.1 for the mechanical-intruder experiments. In (d) the dotted line corresponds to the equation for 2D quasistatic silos \cite{melo_drawbody_2007}.}
\label{fig:all_dpts}
\end{figure*}

We first analyze the long-term displacements and compare the two methodologies. The long-term vertical displacements inside the packing (predominant to horizontal ones due to gravity) are shown in Fig.~\ref{fig:all_dpts}(a) for the melting block and Fig.~\ref{fig:all_dpts}(b) for the mechanical intruder. These results are obtained by averaging the total vertical displacement fields over each set of experiments. As the main reorganization effects are induced by gravity, static zones are observed below the intruder and we thus choose to set the intruders at the bottom of the cell. Both vertical displacement fields share strong qualitative similarities even if the amplitudes are different. Indeed, in a melting block experiment, because of the quasiradial shrinking of the intruder, the maximum vertical displacement is limited to 6.6~$d_g$ instead of 10~$d_g$ that is the intruder's stroke in a mechanical-intruder experiment. However, normalizing the amplitude of the displacements by the maximum amplitude leads to a collapse of both experiments as depicted in Fig. ~\ref{fig:all_dpts}(c) along the vertical axis at $x$~=~0. As a consequence long range displacements are qualitatively equivalent for both experiments and the mechanical intruder is a good method to characterize the displacement due to a shrinking grain although the amplitudes of some radial components are reduced.


In addition, the two displacement fields present similarities with velocity profiles obtained in bidimensional silos when a hole replaces the intruder. However in this study, the amplitude of displacements of the grains is limited by the shape and the geometrical evolution of the intruder and not by free fall dynamics as observed in silos. Our setups are indeed closer to experiments performed for block caving applications \cite{melo_kinematic_2008}. In that study, the authors suggest an analytical formulation for the vertical displacement field in 2D quasistatic silolike systems:
\begin{equation}
\delta(x,y)=-\frac{\delta_0}{2}\left[{\rm erf}\left(\frac{x+R}{\sqrt{4D_{p}y}}\right)-{\rm erf}\left(\frac{x-R}{\sqrt{4D_{p}y}}\right)\right],
\label{eq:dpt_field}
\end{equation}
where $x$ and $y$ are the horizontal and vertical coordinates, respectively; $\delta_0$ is the vertical displacement at the aperture, $R$ is the half width of the aperture; and finally $D_p$ is the diffusion coefficient that links the vertical to horizontal displacements. For the calculation, $R$ is set equal to 20~mm, the radius of the intruder, and $\delta_0$ to the maximum displacement at the edge of the intruder. The only fitting  parameter is $D_p$. In our case, this parameter is found to be close to 6~$d_g$ for both systems. The diffusion coefficient $D_p$ shows large discrepancies in the literature and is usually closer to $d_g$ \cite{choi_velocity_2005}, and eventual dependence on the system parameters is not described. For instance, the geometry of the intruder could play a role. Vertical displacements experimentally found along several horizontal axes are compared to the Eq. \eqref{eq:dpt_field} in Fig.~\ref{fig:all_dpts}(d) for the mechanical intruder. The agreement is again very good for the shrinking intruder as soon as $\delta_0$ is properly set. In summary, these similarities imply a silolike behavior of our experiments for long-term displacements but give no information on the short-term dynamic. In the next section, we focus on the instantaneous responses of the granular packing. 


\section{Short-term displacements}  \label{section:short_term_dpts}

\begin{figure}
\centering
\includegraphics{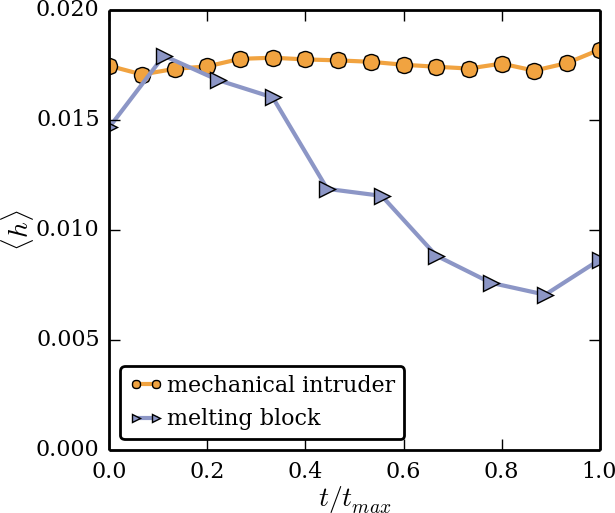}
\caption{Evolution of the dimensionless mean displacement $\langle h \rangle$ averaged over all disks as a function of the normalized time $t/t_{max}$ for the melting block (blue triangles) and the mechanical-intruder (orange circles) experiments. Note that the typical standard deviation on each orange circle is equal to $4.6 \times 10^{-3}$. }
\label{fig:mndpt}
\end{figure}

Because of several experimental limitations mentioned in Sec. \ref{section:experiments}, melting block experiments are not conducive to reliable statistics. Besides, as the size of the intruder is evolving in time, the response of the granular medium is time-dependent and the averaged displacements decrease with time (Fig.~\ref{fig:mndpt}). Indeed, reducing the block size below a few $d_g$ increases the probability for the surrounding structure to jam. Therefore, instantaneous responses at different moments cannot be compared quantitatively. Therefore, we mostly analyzed the short-term displacements using the mechanical-intruder experiment and some qualitative comparisons are established with the melting block experiment.

\subsection{Avalanches}

\begin{figure}
\centering
\includegraphics{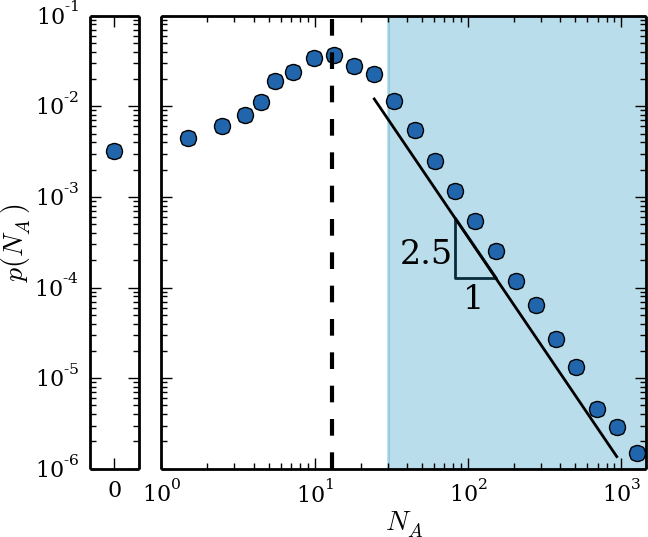}
\caption{Probability density function of the avalanche sizes for 16000 events with the mechanical intruder. The frequency distribution has a slope of -2.5 and the dotted vertical line corresponds to the number of disks in contact with the top of the intruder. The blue region shows the events used for the analysis performed in Fig.~\ref{fig:waiting_time}.  }
\label{fig:all_ava_size}
\end{figure}

\begin{figure}
\centering
\includegraphics{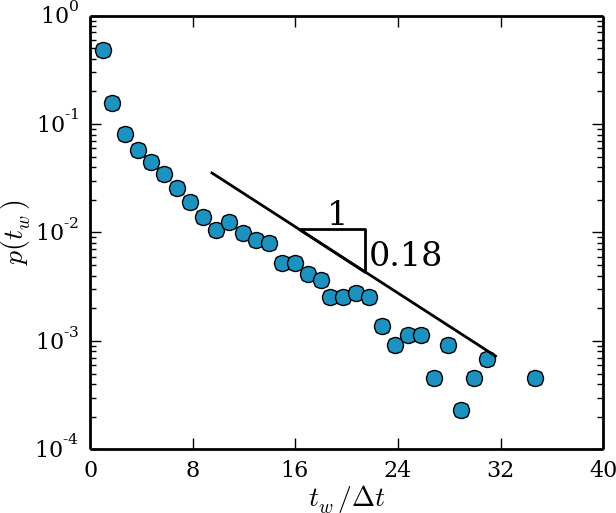}
\caption{Probability density function associated with the waiting durations between two events (only events with $N_A \geq 30$ are considered) normalized by the time step $\Delta t = 5\,{\rm s}$ for the mechanical-intruder experiments.}
\label{fig:waiting_time}
\end{figure}

The instantaneous response of the granular packing is analyzed for the mechanical-intruder experiment. An avalanche of size $N_A$ is defined as the number of disks that have an absolute displacement larger than the displacement of the intruder between two consecutive snapshots ($d_{g}/16$). No phenomenological differences on displacements have been found between the beginning and the end of a mechanical-intruder experiment as shown by the dimensionless mean displacement $\langle h \rangle$ averaged over all disks between two consecutive snapshots with respect to the normalized time in Fig.~\ref{fig:mndpt}. Consequently, the avalanche sizes for the 100 mechanical-intruder experiments have been gathered on a single distribution in Fig.~\ref{fig:all_ava_size}. The distribution spans over three decades, up to large ones involving about one third of the total number of disks. This distribution is characteristic of the length scales and the granular nature of the system. First, there is a nonzero probability to have $N_A=0$ that can be assimilated to quasijamming events even for an intruder as wide as 10~$d_g$. Quasijamming events cannot be assimilated to purely jamming events because the amplitudes of displacements are never strictly null all over the packing. Second, the most frequent avalanche size corresponds to the mean number of disks in contact with the intruder as shown by the dotted vertical line in Fig.~\ref{fig:all_ava_size}. Most of the time, only disks touching the intruder are involved in the avalanche and the perturbation induced by the intruder is damped by frictions or geometrical frustrations over grains. Third, largest events are distributed following a power law. This behavior is characteristic of processes ruled by scale invariance like snow avalanches \cite{birkeland_power-laws_2002} and earthquakes \cite{gutenberg_magnitude_1956}. Note that this property is independent of the threshold used to define avalanches. For the system size and the chosen threshold, a slope of the power-law equal to $-2.5$ captures the behavior of the large events. Although slightly different, this result is qualitatively consistent with other avalanche like systems, in which the slope of the probability of occurrence of largest events is usually found between $-1$ and $-2$ \cite{ramos_avalanche_2009,richter_dynamics_2012,bares_fluctuations_2014}.

In avalanche like systems, the predictability of largest events is a crucial point \cite{ramos_avalanche_2009}. The distribution of the waiting times between two large events is shown in Fig.~\ref{fig:waiting_time}. Here, the largest avalanches have been arbitrarily defined as $N_A \geq 30$. This threshold has been chosen for being in the power-law distribution of avalanche sizes and keeping a good probability of occurrence (Fig.~\ref{fig:all_ava_size}). Another cutoff in the power law would provide the same qualitative results. For correlated events, a power-law distribution of waiting times is expected \cite{vandewalle_avalanches_2001}. However, the largest avalanches exhibit an exponential decay as shown in Fig.~\ref{fig:waiting_time}, with a characteristic time of occurrence $\tau$ of five time steps for the threshold chosen ($\tau \sim 5\,\delta t$):
\begin{equation}
p(t_w) \propto \exp(-t_w \ / \ \tau)
\label{eq:waitingtime}
\end{equation}
This exponential decay can be related to uncorrelated events in time.
For short waiting times, a steeper slope is observed and can be attributed to time clustering of avalanches. This corresponds to quasiconsecutive large events assimilated to foreshocks and aftershocks even if the sequence of events seems unpredictable \cite{ramos_avalanche_2009}. 


To conclude, the short-term displacements share similarities with avalanche like systems, {\textit{i.e.}}, power-law distributed events uncorrelated in time. In the following part, the structural response is analyzed with respect to the event amplitudes.

\subsection{Structural evolution}

\begin{figure}
\centering
\includegraphics{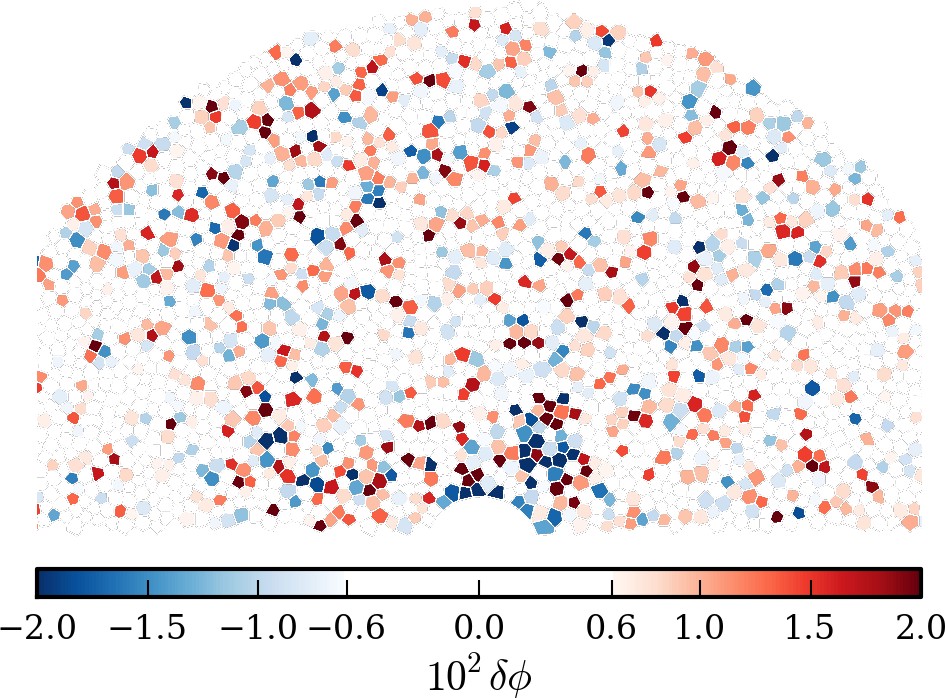}
\caption{Solid fraction variations at the grain scale following a rearrangement event of $\langle h \rangle = 0.018$ in a mechanical-intruder experiment. The range of solid fraction variations has been limited to $-2 \times 10^{-2}$ and $2 \times 10^{-2}$. Being under the accuracy of the measurement, grains with solid fraction variations smaller than $6 \times 10^{-3}$ appear in white. Darkest regions correspond to largest amplitudes of solid fraction variations.}
\label{fig:example_structure}
\end{figure}

Dynamical rearrangements entail microstructural reorganizations at different spatial scales. In this part, the evolution of the granular structure during an event is discussed. To sort the events in order of increasing amplitudes in the rearrangement process, we measure the mean displacement $\langle h \rangle$ averaged over all disks between two consecutive snapshots, which is less arbitrary than the avalanche size. Solid fractions at disk scales are measured using Voronoi tessellations \cite{okabe_spatial_2009}. Such a construction defines a unique spatial division that allocates a local volume to each disk, thus defining the solid fraction at the scale of disks. The local densities are tracked over time and the corresponding local solid fraction variations are then measured between two consecutive snapshots with an accuracy of $6 \times 10^{-3}$. The forthcoming analysis has been performed on a subdomain, in restricted populations of grains, with different geometries to focus on relevant data: the rectangular frame depicted in Fig.~\ref{fig:all_dpts}(a) or the radial geometry in Fig.~\ref{fig:example_structure}, for example. 


A spatial view of the local solid fraction variations is plotted in Fig.~\ref{fig:example_structure} for an event corresponding to $\langle h \rangle = 0.018$. Significant amplitudes of compaction ($\delta\phi >0$) and decompaction ($\delta\phi <0$) occur all over the system and not only close to the intruder. This observation indicates that the granular medium as a whole is involved in the reorganization process even for a localized perturbation. Moreover, the regions undergoing compaction and decompaction are not isolated, but rather interconnected.

\begin{figure}
\centering
\includegraphics{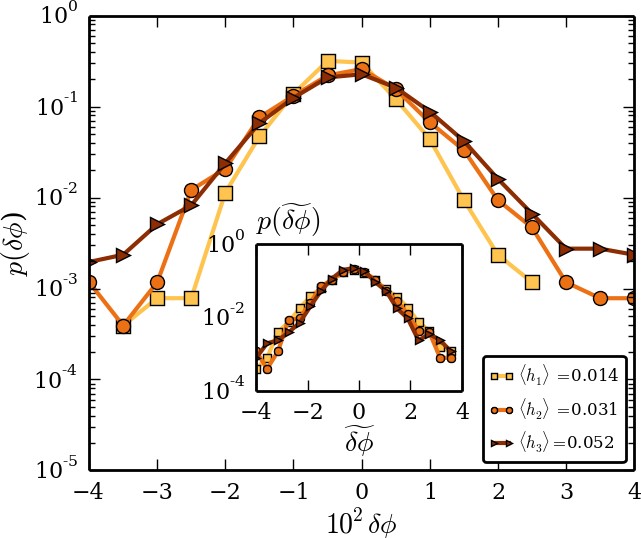}
\caption{Probability density function of local solid fraction variations for three events of different amplitudes in mechanical-intruder experiments: $\langle h_{1} \rangle~=~0.014$ (yellow squares), $\langle h_{2} \rangle~=~0.031$ (orange circles) and $\langle h_{3} \rangle~=~0.052$ (brown triangles). Inset: Probability density function of rescaled local solid fraction variations.}
\label{fig:exp_loc_comp_decomp}
\end{figure}

The probability density function of the local solid fraction variations $p(\delta\phi)$ is displayed in Fig.~\ref{fig:exp_loc_comp_decomp} for three events with different amplitudes ($\langle h_{1} \rangle~=~0.014$, $\langle h_{2} \rangle~=~0.031$, and $\langle h_{3} \rangle~=~0.052$). These distributions are almost centered on $\delta\phi=0$ implying that most of the disks do not undergo strong structural variations. Then, large local compactions always go along with large local decompactions, which is consistent with Fig.~\ref{fig:example_structure}. An explanation could be that the packing needs to create voids somewhere to unjam elsewhere. Finally as the amplitude of events decreases, the local solid fraction variation distribution gets narrower. This result suggests that there is a relation between the structural evolution and the amplitude of events. 

\begin{figure}
\centering
\includegraphics{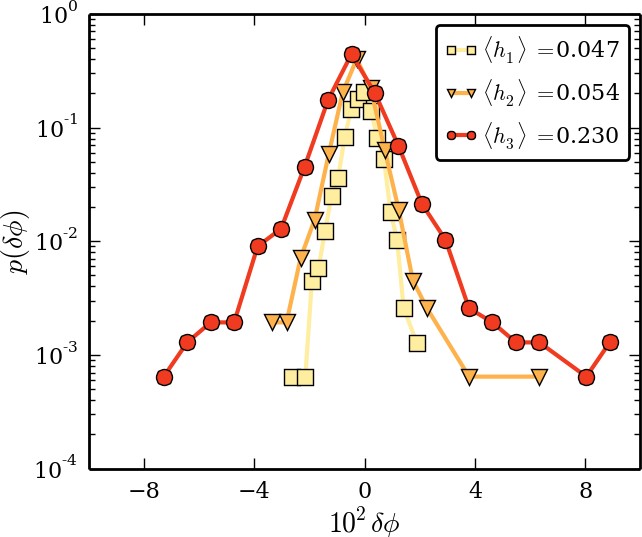}
\caption{Probability density functions associated to the variations of the local solid fractions for three different events showing different mean displacements in melting block experiments. The average variations of solid fractions are $\langle\delta\phi_1\rangle=-2\times 10^{-4}$ (yellow squares), $\langle\delta\phi_2\rangle=-10^{-4}$ (orange triangles) and $\langle\delta\phi_3\rangle=6\times 10^{-4}$ (red circles).}
\label{fig:delta_phi}
\end{figure}

The same conclusions can be obtained in the melting block experiment, as shown in Fig.~\ref{fig:delta_phi}. We note that the amplitudes of displacements are different because the populations considered are not the same. Indeed, we select a smaller population of grains for the melting block experiment due to the smaller size of cell and boundary effects. However, the probability density functions of the local solid fraction variations are quantitatively close for similar amplitudes of displacements in both experiments. We then conclude that the structural evolutions observed in the mechanical-intruder experiment are very similar to the ones observed in the melting block experiment. In the following, we expand on the results obtained with the mechanical intruder rather than on the ones of the too sensitive experiment of the melting block.

\begin{figure}
\centering
\includegraphics{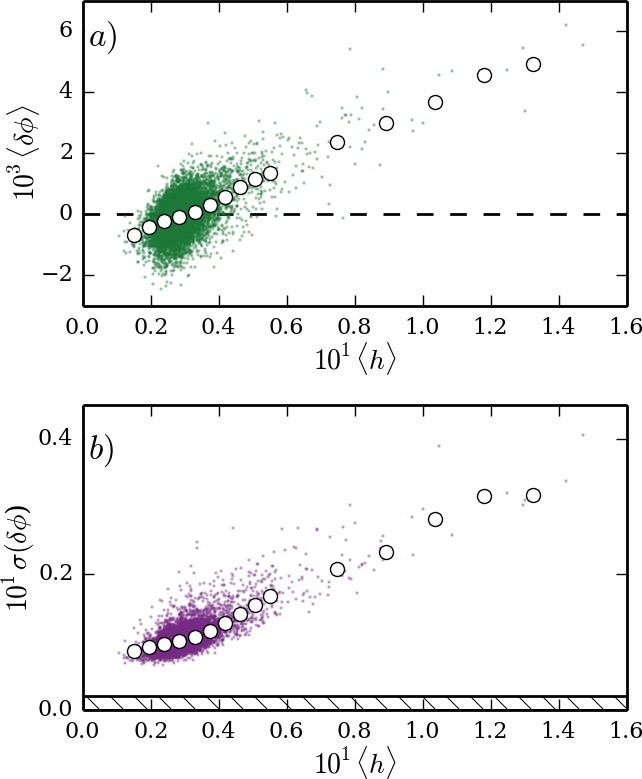}
\caption{(a) Mean value $\langle\delta\phi\rangle$ and (b) standard deviations $\sigma(\delta\phi)$ of local solid fraction variations as a function of the mean displacements $\langle h \rangle$ in mechanical-intruder experiments. White circles correspond to averaged values. In (b) the hashed area corresponds to the limit of detection.}
\label{fig:std_mean_var}
\end{figure}

In the following, statistical informations on the distribution of the solid fraction variations [average $\langle\delta\phi\rangle$ and standard deviation $\sigma(\delta\phi)$] is plotted with respect to the mean displacements: the average $\langle\delta\phi\rangle$ in Fig.~\ref{fig:std_mean_var}(a) and the standard deviation $\sigma(\delta\phi)$ in Fig. \ref{fig:std_mean_var}(b) respectively. All events associated to the mechanical-intruder experiments are plotted in this figure and so correspond to a reliable statistics of about $16\,000$ points. In Fig.~\ref{fig:std_mean_var}, the points are dense around small values of $\langle h \rangle$, in agreement with Fig.~\ref{fig:all_ava_size}, where small avalanches are very frequent. The selected population of grains (the same as in Fig.~\ref{fig:example_structure}) shows higher mean displacements than in the overall packing (Fig.~\ref{fig:mndpt}). Looking at moving averages in Fig.~\ref{fig:std_mean_var}, several qualitative features can be deduced. First, the average solid fraction variations close to zero correspond to short-range mean displacements. Low values of solid fraction variations are associated with small displacements in the packing. Second, a high level of compactions is associated with high average displacements. To get denser, the medium starts from a looser state that allows large amplitude displacements. Third, the moving average of standard deviations monotonically increases with $\langle h \rangle$. Since each distribution is mainly centered around zero, higher standard deviations necessarily correspond to higher amplitudes of local solid fraction variations and so wider distributions. Knowing the link between event amplitudes and parameters of local solid fraction variations distributions, we center and normalize the previous distributions in the inset of Fig.~\ref{fig:exp_loc_comp_decomp} by their average $\langle\delta\phi\rangle$ and standard deviation $\sigma(\delta\phi)$. The three curves collapse on the same distribution. Therefore events with different amplitudes share a common rearrangement mechanism that leads to self-similar statistics of solid fraction variations. This common rearrangement mechanism is described by two parameters: $\langle\delta\phi\rangle$, the mean value of local solid fraction variations basically being the outcome response, and $\sigma(\phi)$, the standard deviation of local solid fraction variations.

\begin{figure}
\centering
\includegraphics{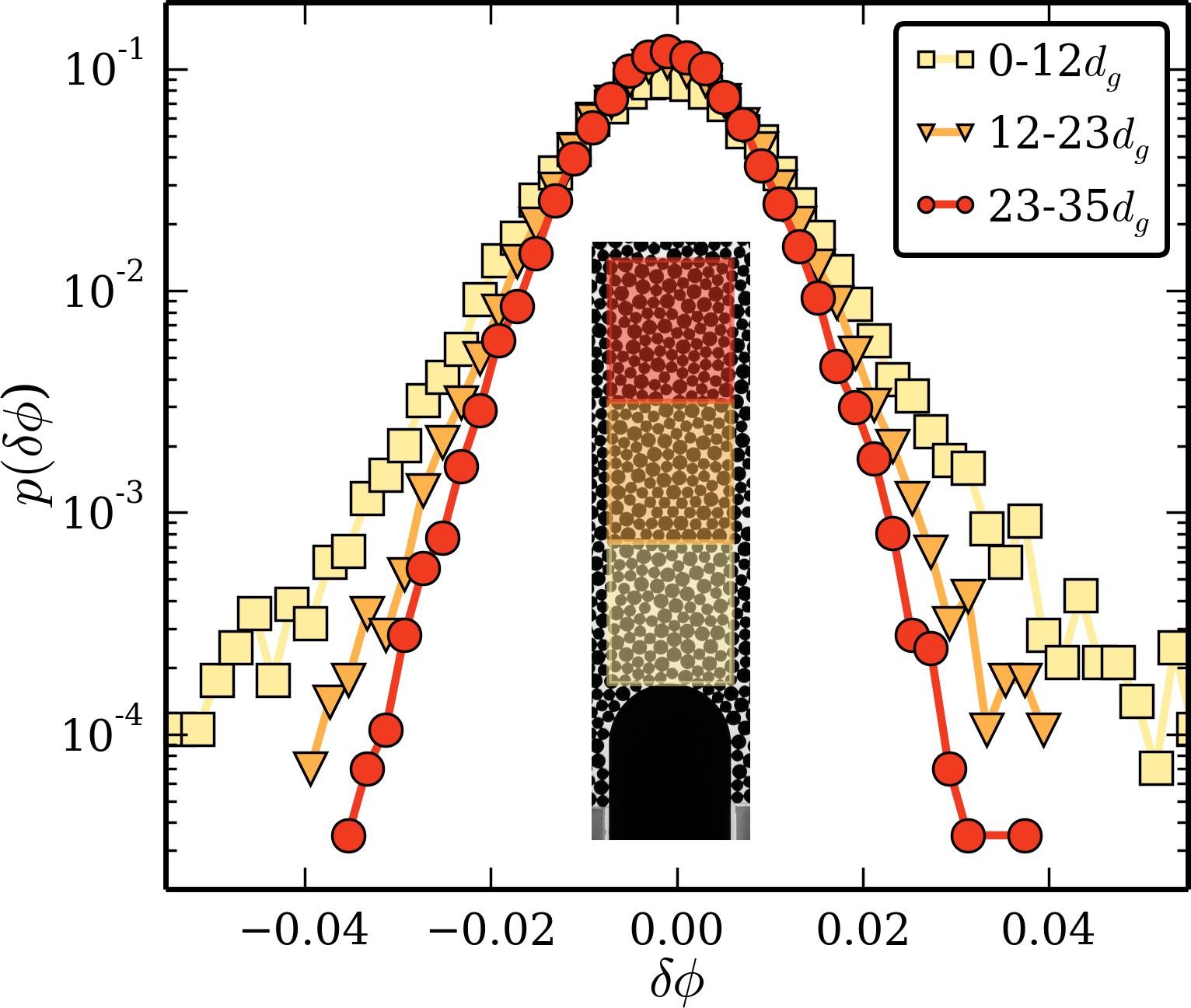}
\caption{Probability density functions of local solid fraction variations in different regions on top of the intruder (see inset) in mechanical-intruder experiments.}
\label{fig:loc_comp_decomp_column}
\end{figure}

\begin{figure}
\centering
\includegraphics{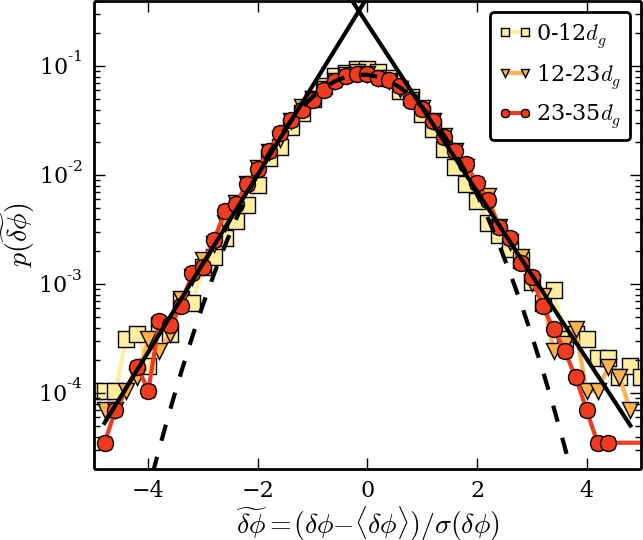}
\caption{Probability density functions of normalized distributions of Fig.~\ref{fig:loc_comp_decomp_column} in mechanical-intruder experiments. The dashed line corresponds to a Gaussian fit over all the distribution. The continuous lines correspond to exponential fits calculated over values of $\widetilde{\delta\phi}$ in [-5,-1] and [1,5].}
\label{fig:loc_comp_decomp_ogeo}
\end{figure}

We ensured that this result does not depend on the position of disks. Indeed, the same analysis is performed for three populations of grains at different distances from the intruder. Given the number of disks in each population, the distribution of the solid fraction variation is measured over three experiments to obtain sufficient statistics. These distributions are shown in Fig.~\ref{fig:loc_comp_decomp_column} and colored according to their vertical positions in the packing. Normalized distributions also collapse (Fig.~\ref{fig:loc_comp_decomp_ogeo}). Choosing a different geometry or another location does not change the conclusions supporting the universality of structural rearrangements in a granular medium undergoing a localized transformation. 

\section{Discussion}  \label{section:discussion}

\begin{figure}
\centering
\includegraphics{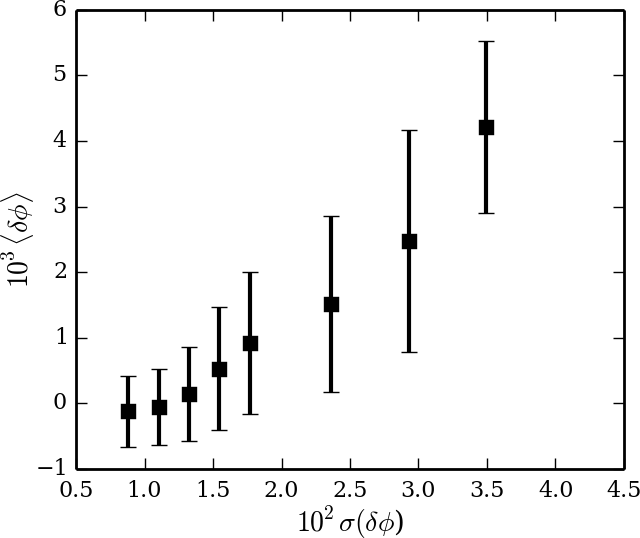}
\caption{Averaged local solid fraction variations in mechanical-intruder experiments as a function of the corresponding standard deviations using the data of Fig.~\ref{fig:std_mean_var}.}
\label{fig:stdmn}
\end{figure}

In this part, the underlying mechanism leading to these distributions of local solid fraction variations is discussed. The forms of the distributions can give clues about the mechanism involved (\textit{e.g.}, \cite{radjai_turbulentlike_2002}). In our study, the shapes of the distributions are conserved independently of the event amplitude. Those distributions are then fitted with Gaussian and exponential curves on Fig.~\ref{fig:loc_comp_decomp_ogeo}. A Gaussian fit does not capture the tails on both sides of the solid fraction variation distributions whereas exponential fit does. Other studies on two-dimensional silo discharges \cite{choi_diffusion_2004,arevalo_anomalous_2007} show similar probability density functions with fat tails on relative displacement fluctuations as long as mean displacements are lower than a particle diameter. Interestingly, this threshold is found to set the transition from a superdiffusive to a diffusive regime. Even if the fluctuations of displacements and the variations of the solid fraction at the grain scale are not strictly equivalent, they still depict the same phenomenon: indeed a small fluctuation of relative displacements induces a local density variation. Such shapes are found in our study for mean displacements $\langle h \rangle$ between 0.03 and 0.2 that are consistent with the superdiffusive regime found in both references. In addition, it is worth noting that the previous references deal with flowing grains whereas we analyze quasistatic experiments. Looking at the exponential tails, we propose a scenario that can give some explanations about the evolution of solid fraction variations. During an event, each disk has a given probability to reach a characteristic value of compaction or decompaction. This value sets the range of solid fraction variations and certainly depends on disk properties. But during this process of reorganization, all the disks evolve until they suddenly stop at a random time which does not depend on either the elapsed time or the amplitude of solid fraction variations reached. Such a scenario would explain the tendencies observed in Fig \ref{fig:loc_comp_decomp_ogeo}. Indeed, exponential tails on both sides indicate a memoryless process for which a uniform and random process is involved.

In addition, structural variations of the packing are found to be correlated to the amplitudes of displacements as depicted in Fig.~\ref{fig:std_mean_var}. Interestingly, in compaction experiments \cite{nicolas_compaction_2000}, similar conclusions can be found linking the amplitude of the shear perturbation to the amplitude of compaction at the scale of the packing. Our results show that only two parameters are needed to entirely describe the probability density functions of local density variations during an event: the mean and the standard deviation of the solid fraction variations. Moreover, this description could be done with only one parameter, since $\sigma(\delta\phi)$ and $\langle\delta\phi\rangle$ appear to be strongly correlated as illustrated in Fig.~\ref{fig:stdmn}.

The circular geometries have been chosen to induce radial solicitations and may promote the formation of arches. It is worth noting that the mechanical-intruder experiment does not fully reproduce the lateral influence induced by a pure radial melting like in a melting block experiment. However, because of the gravity, the vertical motions remain predominant in the rearrangement process, reducing the phenomenological differences between both experiments. Moreover, as depicted in Figs.~\ref{fig:exp_loc_comp_decomp} and \ref{fig:delta_phi}, quantitative similarities are found between the structural evolutions of the packing between both experiments. However, some differences still remain. As depicted in Fig.~\ref{fig:mndpt}, the characteristic size of the melting block plays an important role in the reorganization process. The probability to get large mean displacements decreases with the intruder size, as observed in silos, since the number of disks in contact with the block decreases. In a mechanical-intruder experiment, the mean number of disks in contact with the top disk-shaped region of the intruder remains constant, so this effect does not appear. In addition, the melting of a block within the granular structure remains intrinsically closer to real processes than a mechanical intruder. But such experiments are still too sensitive to get reliable statistics, that is why we decided to expand our results on mechanical-intruder experiments that are more accurate. Therefore, numerical simulations could benefit this study to explore the process of rearrangements with different dynamics of melting or with higher numbers of melting grains with a better repeatability.


In conclusion, we have studied experimentally the rearrangements of a granular medium undergoing a localized transformation. Mainly driven by gravity, the granular systems exhibit long-term displacements similar to those from 2D quasistatic silos. In the mechanical-intruder experiment, local instantaneous displacements are analyzed statistically through the concept of avalanches whose sizes range over three decades. The size distribution of the largest events follows a power law, which is characteristic of processes ruled by scale invariance. Moreover, these large events are uncorrelated in time. Finally, all events participating in the structural evolution of the system share a common mechanism of rearrangements independently of their amplitudes. It shows that only two parameters are necessary to fully describe these structural reorganizations: the mean value that sets the outcome response and the standard deviation. We propose that the origin of the underlying mechanism is governed by a homogeneous and random process. 


Most industrial processes involve fully reactive packings. In this study, we only considered a localized transformation due to a single intruder. Even for such a localized perturbation, the granular packing exhibits a long-range response that can involve almost all the packing. Therefore, we expect that multiple and localized transformations, such as a second intruder, may have a significant influence on the packing and exhibit cooperative effects \cite{kunte_spontaneous_2014}.

\subsection*{acknowledgments}

This work has benefited from interesting discussions with P. Levitz and O. Ramos.

\bibliography{bibliography}

    \bibliographystyle{ieeetr}

    \end{document}